\documentclass[sn-mathphys-num]{sn-jnl}

\usepackage{graphicx}%
\usepackage{multirow}%
\usepackage{amsmath,amssymb,amsfonts}%
\usepackage{amsthm}%
\usepackage{mathrsfs}%
\usepackage[title]{appendix}%
\usepackage{xcolor}%
\usepackage{textcomp}%
\usepackage{manyfoot}%
\usepackage{booktabs}%
\usepackage{algorithm}%
\usepackage{algorithmicx}%
\usepackage{algpseudocode}%
\usepackage{listings}%
\usepackage{makecell}

\theoremstyle{thmstyleone}%
\theoremstyle{thmstyletwo}%

\theoremstyle{thmstylethree}%

\def\etal{{\it et al.}}

\begin{document}

\title[NPGPT: NP-like Compound Generation with CLMs]{NPGPT: Natural Product-Like Compound Generation with GPT-based Chemical Language Models}


\author[1]{\fnm{Koh} \sur{Sakano}}\email{sakano@li.c.titech.ac.jp}

\author[1]{\fnm{Kairi} \sur{Furui}}\email{furui@li.c.titech.ac.jp}

\author*[1]{\fnm{Masahito} \sur{Ohue}}\email{ohue@comp.isct.ac.jp}

\affil[1]{\orgdiv{Department of Computer Science, School of Computing}, \orgname{Institute of Science Tokyo}, \orgaddress{\city{Kanagawa}, \country{Japan}}}


\abstract{Natural products are substances produced by organisms in nature and often possess biological activity and structural diversity. Drug development based on natural products has been common for many years. However, the intricate structures of these compounds present challenges in terms of structure determination and synthesis, particularly compared to the efficiency of high-throughput screening of synthetic compounds. In recent years, deep learning-based methods have been applied to the generation of molecules. In this study, we trained chemical language models on a natural product dataset and generated natural product-like compounds. The results showed that the distribution of the compounds generated was similar to that of natural products. We also evaluated the effectiveness of the generated compounds as drug candidates. Our method can be used to explore the vast chemical space and reduce the time and cost of drug discovery of natural products.}




\maketitle

\section{Introduction}
Natural products derived from plants and microorganisms have attracted attention for their beneficial properties and diverse biological activities~\cite{Dias2012-jq, cragg2013natural}. These compounds are known for their complex structures and large molecular weights. Because they are biosynthesized within living organisms, many of them display potent biological activities and are often used as lead compounds in drug development. Between 1981 and 2002, natural products accounted for over 60\% and 75\% of the new chemical entities (NCEs) developed for cancer and infectious diseases, respectively~\cite{Newman2016-pm}. In addition, approximately half of the drugs currently available on the market are derived from natural products~\cite{Demain2014-hl}, highlighting their vital role in drug discovery and development.

The unique molecular structures of natural products, which are rarely found in synthetic compounds, contribute to their biological activity~\cite{Dias2012-jq}. The golden age of natural product drug discovery began in the 1940s with the discovery of penicillin. Many drugs were discovered from microbes, especially actinomycetes and fungi, until the early 1970s. However, from the late 1980s to early 1990s, new drug discoveries from natural products declined~\cite{pelaez2006historical}. Pharmaceutical companies began to withdraw from natural product research due to the emergence of combinatorial chemistry and high-throughput screening (HTS), which allowed artificial creation of chemical diversity. Furthermore, the complexity of the structures of natural products made synthesis and derivatization difficult, complicating lead compound optimization~\cite{shen2015new, li2009drug}. Despite these challenges, natural products have recently been reassessed and are once again gaining attention as valuable resources in drug discovery due to their diverse structures and biological activities~\cite{pelaez2006historical}.

In recent years, advances in deep learning-based molecular generation have been used for the discovery of novel pharmaceuticals~\cite{Bilodeau2022-xx}. This approach involves the virtual generation of compounds on computers, with the aim of identifying useful candidate molecules. However, because the training process typically utilizes general chemical databases comprising relatively small molecules such as PubChem~\cite{PubChem2023}, there are challenges to generate large and complex compounds similar to natural products. Consequently, this limitation narrows the chemical space that can be explored~\cite{pmlr-v119-jin20a}.

In this study, we propose a molecular generative model capable of producing natural product-like compounds. By generating a group of molecules using a model that has learned the distribution of natural products, we aim to facilitate the search for lead molecules in drug discovery and reduce the costs of natural product-based drug development.

A closely related previous study to this research is the work of Tay \etal, who used a recurrent neural network (RNN) to generate natural products~\cite{Tay2023-nr}. They trained an RNN equipped with LSTM units on natural products from the COCONUT database~\cite{Sorokina2021-ct} and developed a model capable of generating compounds similar to natural products. They showed that the distribution of natural product-likeness scores of the compounds generated was similar to that of the natural products in COCONUT. This study aims to create a more high-performance model using Transformers compared to the approach of Tay \etal, and further evaluates whether the generated library is useful as a candidate for pharmaceuticals.

\section{Methods}
\subsection{Fine-tuning and chemical language models}
\begin{figure}[tbp]
    \centering
    \includegraphics[width=1\linewidth]{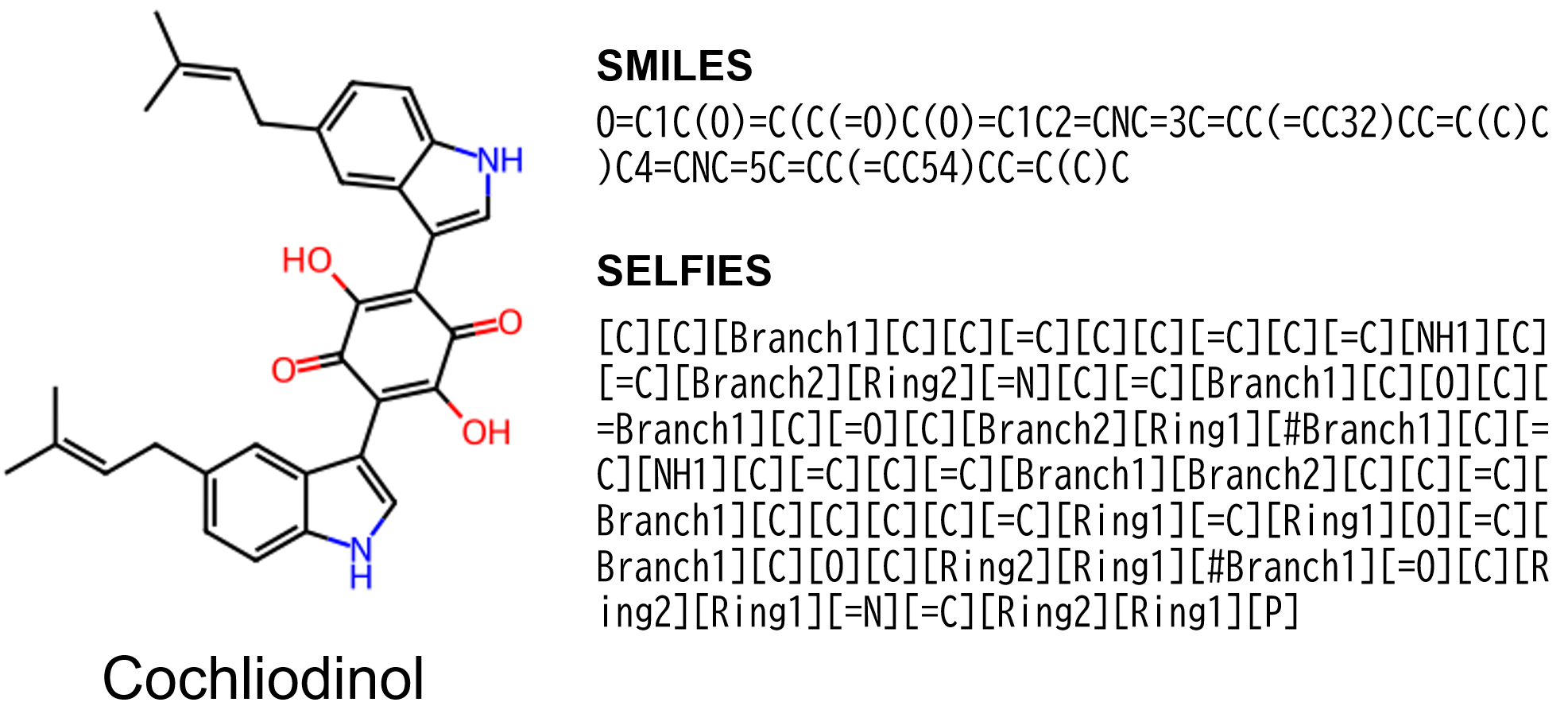}
    \caption{Examples of SMILES and SELFIES encoding (cochliodinol, a natural product compound)}
    \label{fig:f1}
\end{figure}

Fine-tuning language models is a technique that refines models, initially trained on extensive datasets, to excel in particular tasks, tailoring them to specialized requirements. In this study, we fine-tuned chemical language models using a natural product dataset. A chemical language model refers to a model that processes string representations of molecules, e.g., simplified molecular-input line-entry system (SMILES)~\cite{weininger1988smiles} and self-referencing embedded strings (SELFIES)~\cite{krenn2020self}. Examples of these string representations are shown in Figure~\ref{fig:f1}. We hypothesized that since pre-trained models have already learned chemical structures, we can efficiently construct a model capable of generating natural product-like compounds.

\subsection{Dataset}
We used the COCONUT database, which encompasses approximately 400,000 natural products~\cite{Sorokina2021-ct}. As a preprocessing step, we standardized the SMILES strings and removed large compounds (with an atom count greater than 150 or more than 10 rings). Subsequently, we employed a technique that enumerates SMILES by randomizing the traversal order of the molecular graph~\cite{bjerrum2017smiles}, augmenting the data by approximately nine times. The final dataset included approximately 3.6 million entries and was used for the fine-tuning process.

\subsection{Models}
We selected pretrained models that satisfy the following criteria:

\begin{itemize}
    \item It has been trained on a dataset of significant size.
    \item It is a decoder-only model that utilizes only the decoder of a transformer architecture.
\end{itemize}

We selected two models, smiles-gpt~\cite{adilov2021generative} and Chem\-GPT~\cite{Frey2023-os}. The details of the models are shown in Table \ref{tab:pretrained_model}. Both models used the PubChem-10M dataset~\cite{chithrananda2020chemberta} for pretraining (smiles-gpt used the first 5 million molecules of PubChem-10M), and their architecture is based on GPT. They differ in the molecular string representation used: smiles-gpt employs SMILES, whereas ChemGPT uses SELFIES.

\begin{table*}[tbp]
    \centering
    \caption{Pretrained models used in this study.}
    \label{tab:pretrained_model}
    \begin{tabular}{ccccc}
        \toprule
        Model & \makecell{Molecular\\representation} & Pretraining dataset & Architecture & \makecell{Number of\\parameters} \\
        \midrule
        smiles-gpt & SMILES & \makecell{PubChem-10M\\(first half)} & GPT-2~\cite{radford2019language} & 24.8M \\ \midrule
        ChemGPT & SELFIES & PubChem-10M & GPT-Neo~\cite{gpt-neo} & 19M \\
        \bottomrule
    \end{tabular}
\end{table*}

\subsection{Training}
We fine-tuned the models on the natural product dataset using the AdamW optimizer~\cite{loshchilov2017decoupled}. The learning rate was set from $5.0 \times 10^{-4}$ to $5.0 \times 10^{-8}$ (using a cosine annealing schedule) for smiles-gpt and $5.0 \times 10^{-5}$ for ChemGPT. The batch size was set to 256 and 32 for smiles-gpt and ChemGPT, respectively. Due to the lengthy nature of SELFIES and their substantial byte size, the use of SELFIES in ChemGPT necessitated a reduction in batch size due to the constraints imposed by GPU memory capacity. This training was conducted on four GeForce RTX 3090 GPUs.

\section{Results and Discussion}
\subsection{Evaluation of generated molecules}
We calculated validity, uniqueness, novelty, internal diversity~\cite{Polykovskiy2020-hd}, and Fr{\'e}chet {ChemNet} Distance (FCD)~\cite{Preuer2018-mn} for the 100 million molecules generated and made public in a previous study~\cite{Tay2023-nr}, as well as for the 100 million molecules generated by fine-tuned ChemGPT and smiles-gpt.

\begin{itemize}
    \item Validity: The ratio of valid molecules to the total number of generated molecules. A valid molecule is one that can be parsed by RDKit~\cite{rdkit}.
    \item Uniqueness: The ratio of unique molecules to the total number of generated molecules.
    \item Novelty: The ratio of molecules that do not exist in the COCONUT database.
    \item Internal diversity: The average pairwise Tanimoto similarity between the generated molecules, calculated using Morgan fingerprints with a radius of 2 and 1024 bits. This metric was calculated using MOSES~\cite{Polykovskiy2020-hd}.
    \item FCD: A metric of the distance between the distribution of generated molecules and that of training dataset. A smaller FCD indicates that the set of generated molecules are closer to the training data distribution.
\end{itemize}

The results are shown in Table \ref{tab:generated_molecules}. smiles-gpt achieved results close to those of a previous study~\cite{Tay2023-nr}. Compared to Tay \etal, the smaller FCD suggests that more compounds similar to natural products were generated, indicating sampling from a smaller chemical space that is better adapted to the space of natural products. In this respect, it has managed to generate compounds more closely resembling natural products than the previous study.

ChemGPT exhibited high validity, which is believed to be due to the use of SELFIES. However, the significantly large FCD indicates that the distribution of natural products was not captured accurately. Although high uniqueness and novelty are numerically positive outcomes, the magnitude of FCD suggests sampling from a broader chemical space, resulting in the generation of compounds that appear to be nearly random.

We measured the FCD of the molecular sets generated by the models before and after fine-tuning. ChemGPT had an FCD of 29.01 while smiles-gpt had an FCD of 6.75. This indicates that the distribution of molecules generated by ChemGPT changed significantly after fine-tuning compared to smiles-gpt. Although Table \ref{tab:generated_molecules} suggests that ChemGPT may not have learned the distribution of COCONUT, at least the distribution of generated molecules is different.

\begin{table*}[tbp]
    \centering
    \caption{Validity, uniqueness, novelty, internal diversity, FCD of the generated set of 100 million molecules.}
    \label{tab:generated_molecules}
    \begin{tabular}{lccccc}
        \toprule
        Model & Validity $\uparrow$ & Unique $\uparrow$ & Novelty $\uparrow$ & \makecell{Internal\\diversity $\uparrow$} & FCD $\downarrow$ \\
        \midrule
        Tay \etal~\cite{Tay2023-nr} & 0.904 & 0.753 & 0.998 & \textbf{0.885} & 1.794 \\
        smiles-gpt (fine-tuned) & 0.903 & 0.663 & 0.996 & 0.873 & \textbf{1.290} \\
        ChemGPT (fine-tuned) & \textbf{0.999} & \textbf{0.939} & \textbf{0.999} & 0.882 & 14.28 \\
        \bottomrule
    \end{tabular}
\end{table*}

\subsection{Visualization of the distribution in physicochemical space of generated molecules}
\begin{figure}[tbp]
    \centering
    \includegraphics[width=0.7\linewidth]{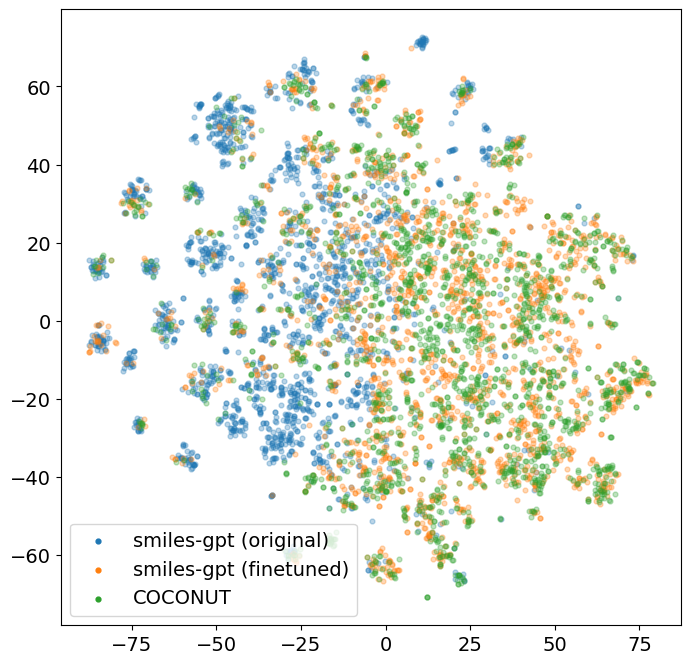}
    \caption{t-SNE visualization of 2,000 molecules generated by the original and fine-tuned models of smiles-gpt, along with molecules from COCONUT.}
    \label{fig:smiles-gpt_tsne}
\end{figure}

\begin{figure}[tbp]
    \centering
    \includegraphics[width=0.7\linewidth]{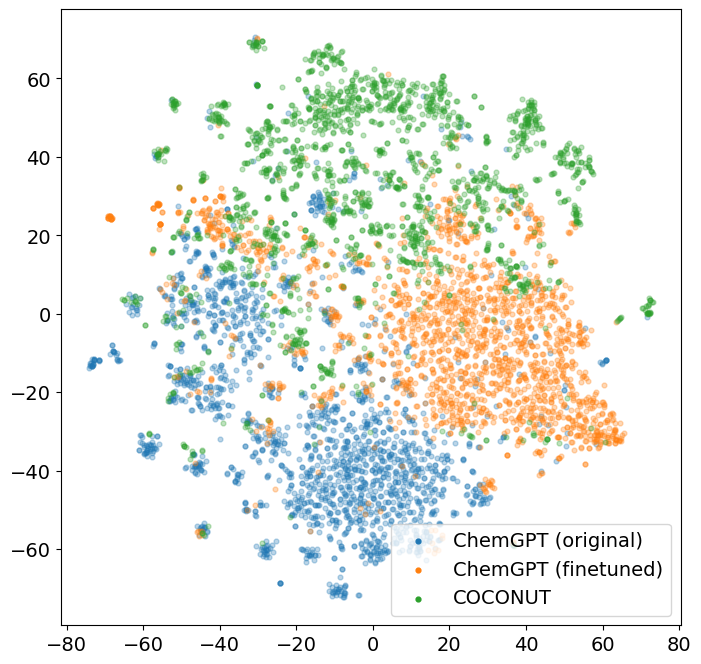}
    \caption{t-SNE visualization of 2,000 molecules generated by the original and fine-tuned models of ChemGPT, along with molecules from COCONUT.}
    \label{fig:chemgpt_tsne}
\end{figure}

We visualized the distribution of molecules generated by the original and fine-tuned models, along with COCONUT compounds, using t-distributed stochastic neighbor embedding (t-SNE). We randomly selected 2,000 molecules from the generated ones and embedded them in two dimensions using t-SNE based on 209 physicochemical descriptors for each molecule. For the calculation of the descriptors, we utilized \texttt{Descriptors.CalcMolDescriptors} from RDKit. The visualization results are shown in Figures \ref{fig:smiles-gpt_tsne} and \ref{fig:chemgpt_tsne}.

From the smiles-gpt results in Figure \ref{fig:smiles-gpt_tsne}, it appears that the overall distribution of the molecules has moved closer to COCONUT through fine-tuning. In contrast, as shown in Figure \ref{fig:chemgpt_tsne}, ChemGPT still exhibits a different distribution from COCONUT even after fine-tuning.

\subsection{Distribution of scores for generated molecules}
\begin{figure}[tbp]
  \centering
  \includegraphics[width=\linewidth]{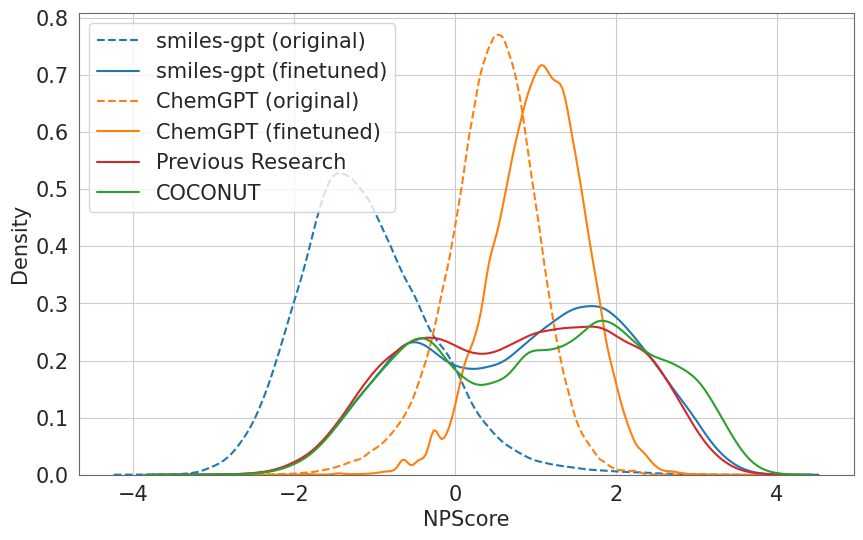}
  \caption{Kernel density estimation of NP Scores for molecules generated by the original and fine-tuned models of smiles-gpt and ChemGPT, compared with molecules generated in the previous research (Tay \etal~\cite{Tay2023-nr}) and natural products from COCONUT.}
  \label{fig:npscore_dist}
\end{figure}

\begin{figure}[tbp]
  \centering
  \includegraphics[width=\linewidth]{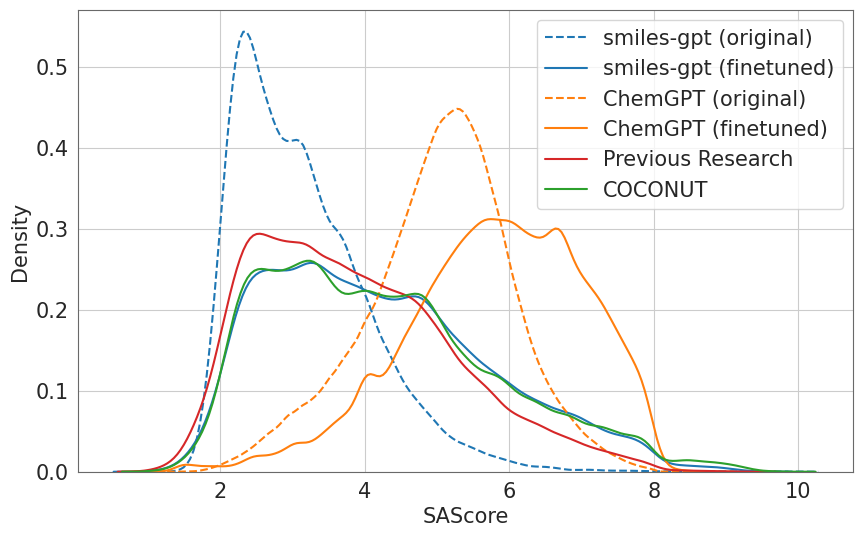}
  \caption{Kernel density estimation of SA Scores for molecules generated by the original and fine-tuned models of smiles-gpt and ChemGPT, compared with molecules generated in the previous research (Tay \etal~\cite{Tay2023-nr}) and natural products from COCONUT.}
  \label{fig:sascore_dist}
\end{figure}

We calculated the natural product-likeness score (NP Score)~\cite{Ertl2008-qc} and the synthetic accessibility score (SA Score)~\cite{Ertl2009-he} for molecules generated by the original model and the model after fine-tuning, as well as for molecules generated in the previous research by Tay \etal, and compared their distributions with those of the natural product data. Kernel density estimation was performed on the NP and SA Scores data for each molecular library, and the results are plotted in Figures \ref{fig:npscore_dist} and \ref{fig:sascore_dist}.

The NP Score is an index that measures the natural product-likeness of a compound, calculated based on the frequency of occurrence of substructures in natural products. The SA Score is an index used to quantitatively assess the synthetic accessibility of a compound, where a lower score indicates a greater ease of synthesis.

smiles-gpt, through fine-tuning, has approached a distribution of both NP Scores and SA Scores closer to those of COCONUT, whereas ChemGPT continues to generate compounds with a significantly different distribution from COCONUT even after fine-tuning. Furthermore, in comparison to previous research, the fine-tuned smiles-gpt is capable of generating compounds that are closer to those in COCONUT, particularly in terms of SA Score.

From the above results, it is evident that fine-tuned smiles-gpt can generate compounds that are more reminiscent of natural products compared to fine-tuned ChemGPT. Although it is difficult to make a definitive statement due to differences in training conditions and model specifics, it is believed that the distinction between SMILES and SELFIES plays a significant role. Although it is advantageous that SELFIES are 100\% valid, they appear to be a more verbose and relatively less intuitive molecular representation compared to SMILES.

Comparative studies between SMILES and SELFIES have reported that SMILES-trained models exhibit better performance~\cite{gao2022sample, ghugare2023searching}. Although the lower validity of SMILES has been a concern, current language models have become sufficiently adept at learning the syntax of SMILES. Gao \etal{} have pointed out that the advantage of SELFIES being 100\% valid is decreasing~\cite{gao2022sample}.

\subsection{Evaluation of bioactivity potential by protein--ligand docking}
The utility of the generated compound library as potential drug candidates was evaluated through protein--ligand docking calculations with proteins. We evaluated the viability of these compounds for pharmaceutical use from protein--ligand interactions.

For the target protein, the epidermal growth factor receptor (EGFR) was selected. Inhibition of EGFR has been reported to significantly suppress cancer cell proliferation~\cite{Normanno2003-wf}, and several EGFR inhibitors have been developed as pharmaceuticals. Gefitinib and erlotinib are among the well-known inhibitor drugs. In this experiment, the crystal structure of EGFR with PDB ID: 2ITY~\cite{Yun2007-yu} was used, which is the complex structure of EGFR and gefitinib.

Initially, 1,000 molecules were randomly selected as ligands from those generated by the fine-tuned smiles-gpt. As indicated in the results above, because the fine-tuned ChemGPT was unable to generate natural product-like compounds, molecules generated by ChemGPT were not used for docking. Subsequently, the ligands were prepared using Schr\"odinger LigPrep~\cite{ligprep}, and the generated 12,930 conformers were docked using Schr\"odinger Glide software version 2020-2~\cite{Friesner2004-vd}.

\begin{figure}[tbp]
    \centering
    \includegraphics[width=1.0\linewidth]{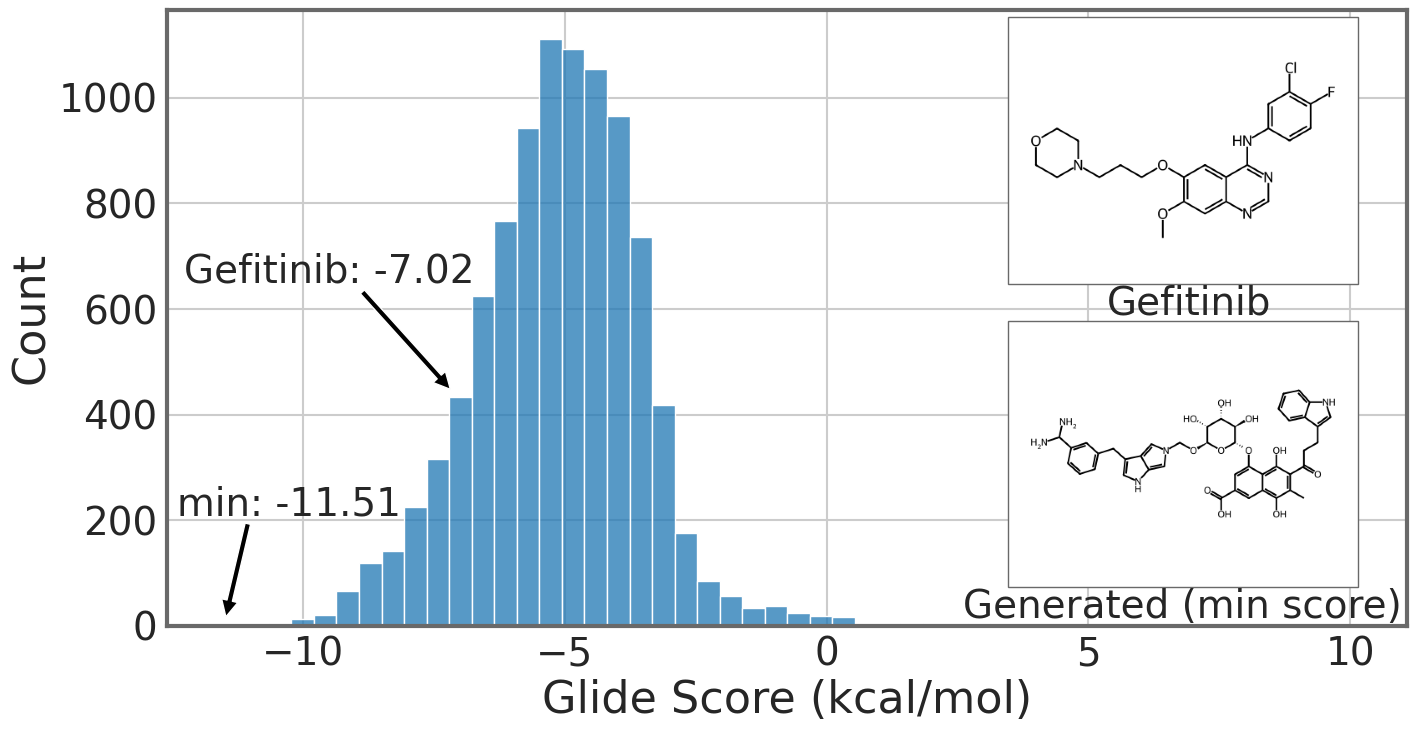}
    \caption{Distribution of GlideScore for 12,930 docked conformations of 1,000 molecules generated by fine-tuned smiles-gpt.}
    \label{fig:glidescore_hist}
\end{figure}

The distribution of GlideScores for each conformation obtained from the docking is shown in Figure~\ref{fig:glidescore_hist}. The GlideScore represents the predicted binding free energy between a protein and a ligand, with lower values indicating stronger binding. Although the GlideScore for gefitinib is $-7.02$ kcal/mol~\cite{Ochiai2023-mj}, there are 1,216 conformations with a better score than gefitinib, accounting for 9.8\% of all docked conformations. Among these, the lowest (best) GlideScore was $-11.51$ kcal/mol. This indicates that a significant number of compounds with docking scores that are better than those of existing inhibitors have been generated.

\begin{figure}[tbp]
    \centering
    \includegraphics[width=1.0\linewidth]{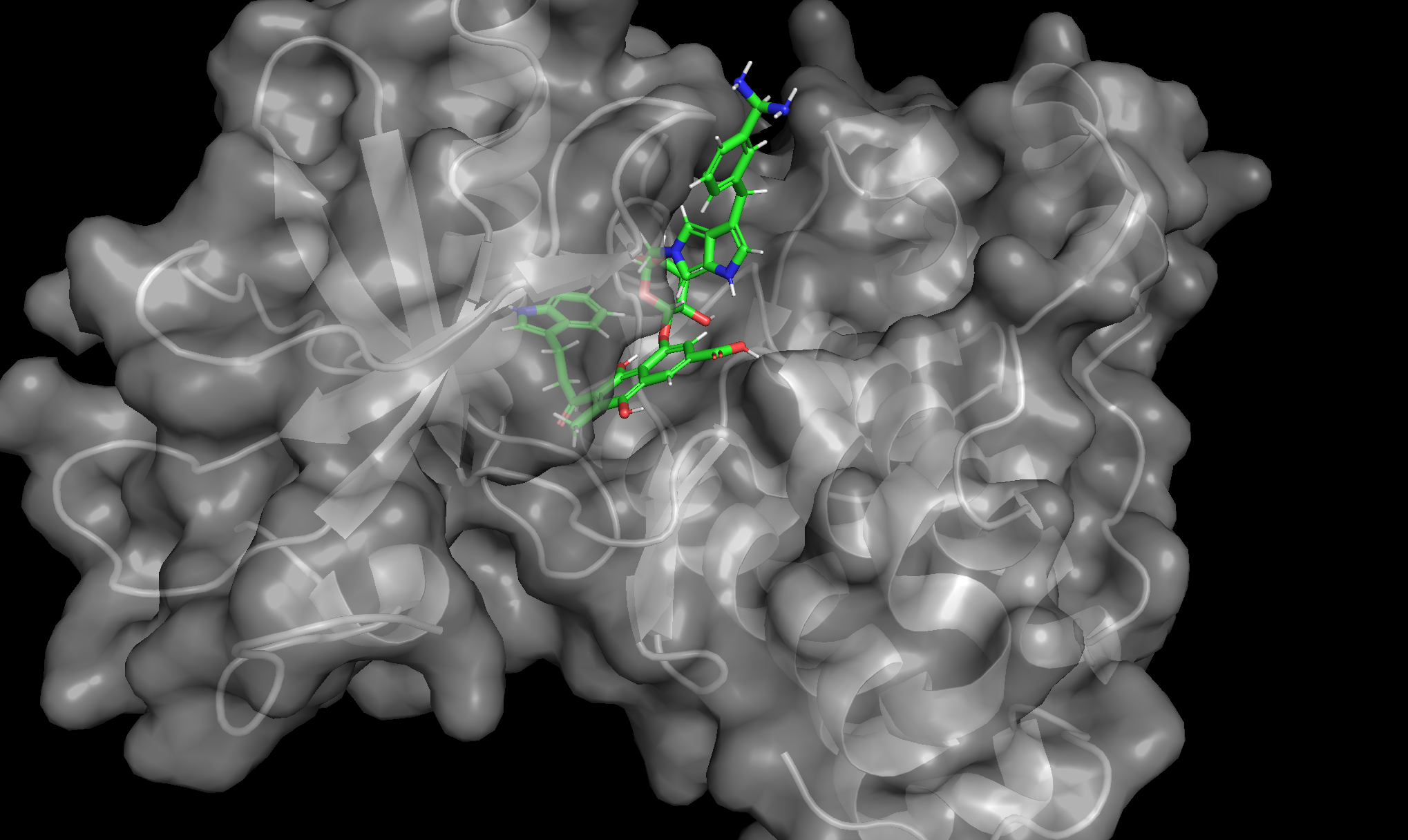}
    \caption{Docking pose of the compound with the best GlideScore with EGFR.}
    \label{fig:pose_top1}
\end{figure}

Table \ref{tab:top10} presents the top 10 compounds with the best GlideScores of the 1,000 compounds subjected to docking, together with the natural products of the COCONUT database that exhibited the highest similarity to each of these compounds. Although compounds with substructures similar to those of gefitinib were generated, most have relatively complex structures. Based on the NP Scores of these 10 compounds, most of them are likely to be natural products, as NP Score ranges from -5 to 5. The SA Scores are relatively high, considering that many compounds in small molecule databases like ChEMBL~\cite{chembl2019} have values of 2-3, suggesting that they are difficult to synthesize. Furthermore, observing similar natural products reveals that the model has successfully learned to build scaffolds of natural products. Figure \ref{fig:pose_top1} shows the docking pose of the compound with the best GlideScore. We can see that the compound is well-adjusted in the EGFR binding pocket.

\begin{figure}[tbp]
    \centering
    \includegraphics[width=0.85\linewidth]{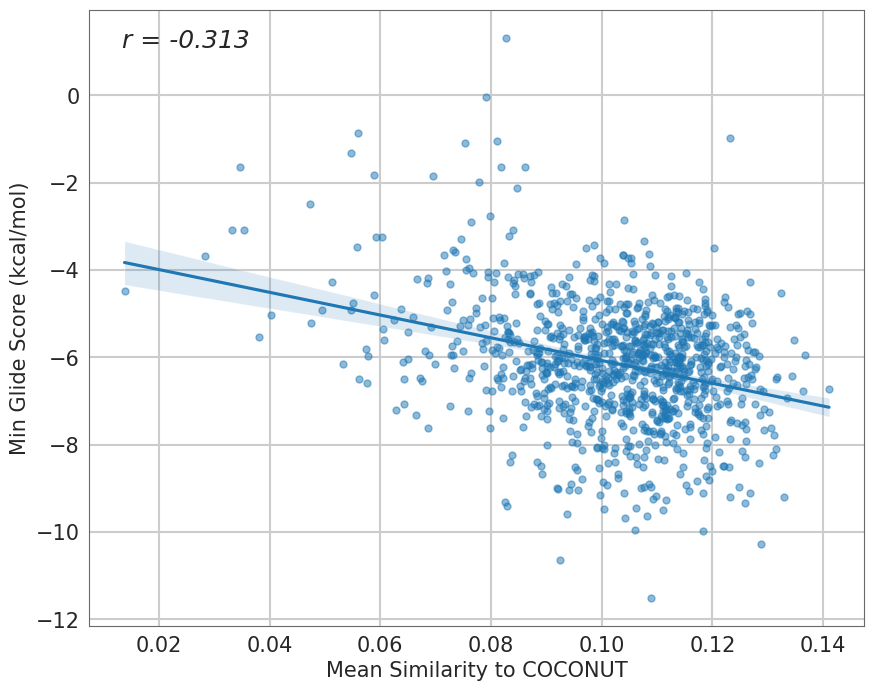}
    \caption{Relationship between the mean similarity with all compounds in COCONUT and GlideScore for 1,000 molecules generated by fine-tuned smiles-gpt.}
    \label{fig:min_glidescore_similarity}
\end{figure}

Furthermore, to verify whether natural product-likeness influences drug-likeness, we calculated the similarity between the natural products and the compounds subjected to docking and investigated the correlation. The Tanimoto index of the Morgan fingerprint with a radius of 2 and 2,048 bits was used for similarity measures. For the 1,000 compounds selected for docking, we calculated their mean similarity to all compounds in the COCONUT database and depicted the relationship with the GlideScore in Figure \ref{fig:min_glidescore_similarity}. Note that the overall similarity is low because they are the mean values. For compounds with multiple stereoisomers, the one with the minimum GlideScore was chosen. The Pearson correlation coefficient between mean similarity to natural products and GlideScore was $r=-0.313$, indicating a modest correlation. Although the difference in the similarity of the generated compounds is slight (0.02 - 0.14), it can be inferred that compounds with a certain degree of natural product-likeness tend to have better docking scores.

\begin{figure}
      \centering
      \includegraphics[width=0.85\linewidth]{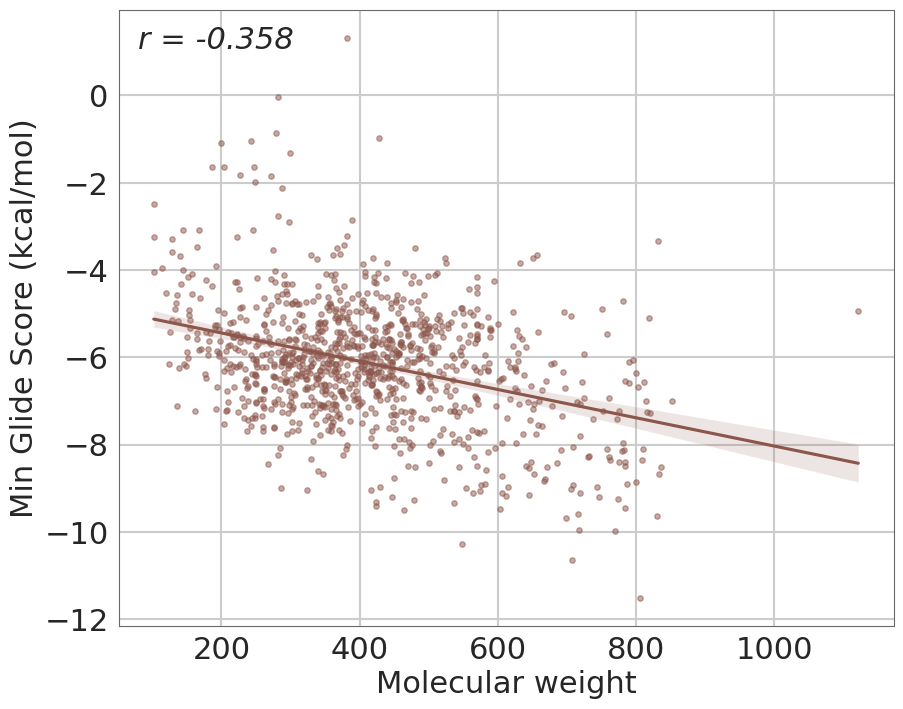}
      \caption{Relationship between molecular weight and GlideScore for 1,000 molecules generated by fine-tuned smiles-gpt.}
      \label{fig:min_glidescore_mw}
\end{figure}

However, it should be noted that there is a tendency for docking scores to improve as the molecular weight increases~\cite{LE1,LE2}. Figure \ref{fig:min_glidescore_mw} shows the relationship between molecular weight and GlideScore for the 1,000 compounds. There is a weak correlation between molecular weight and GlideScore ($r=-0.358$), suggesting that larger molecules tend to have better docking scores. Therefore, the improvement in docking scores cannot be solely attributed to the resemblance to natural products.

\begin{table}[tbp]
\centering
\caption{Structures, natural product-likeness scores and synthetic accessibility scores of the top 10 compounds out of 1,000 compounds, based on GlideScore in the docking experiment, and the natural products with the highest similarity to those compounds}
\label{tab:top10}
\begin{tabular}{cccccc}
\toprule
& Structure & Most similar natural product & \makecell{GlideScore\\(kcal/mol)} & NP Score & SA Score \\
\midrule
1 & \begin{minipage}{8em}
      \centering
      \scalebox{0.11}{\includegraphics{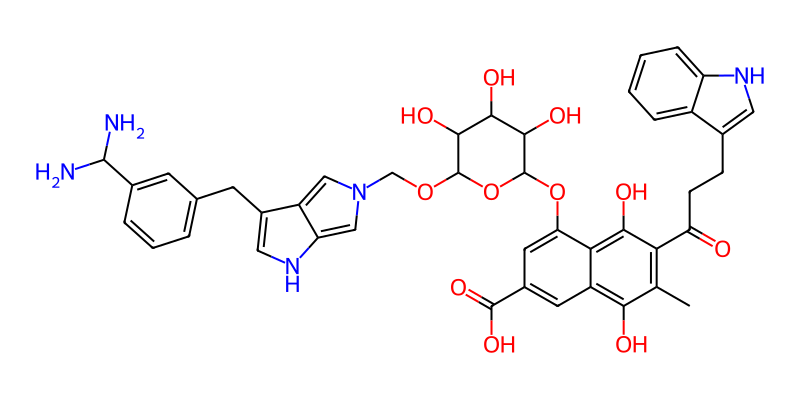}}
\end{minipage} & \begin{minipage}{8em}
      \centering
      \scalebox{0.11}{\includegraphics{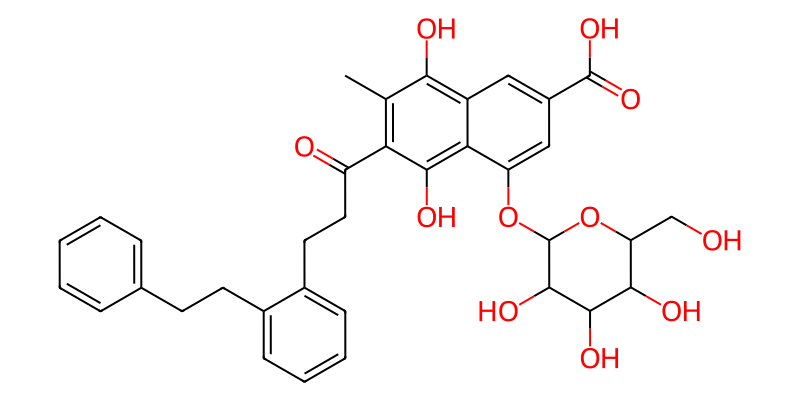}}
\end{minipage} & $-11.51$ & 1.140 & 5.105 \\ \midrule
2 & \begin{minipage}{8em}
      \centering
      \scalebox{0.11}{\includegraphics{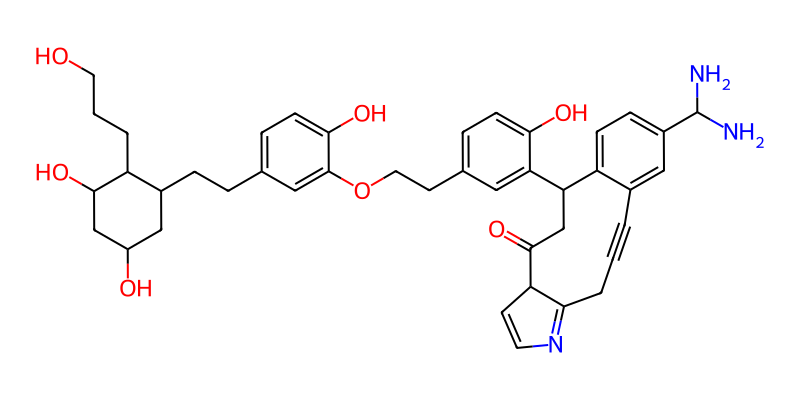}}
\end{minipage} & \begin{minipage}{8em}
      \centering
      \scalebox{0.11}{\includegraphics{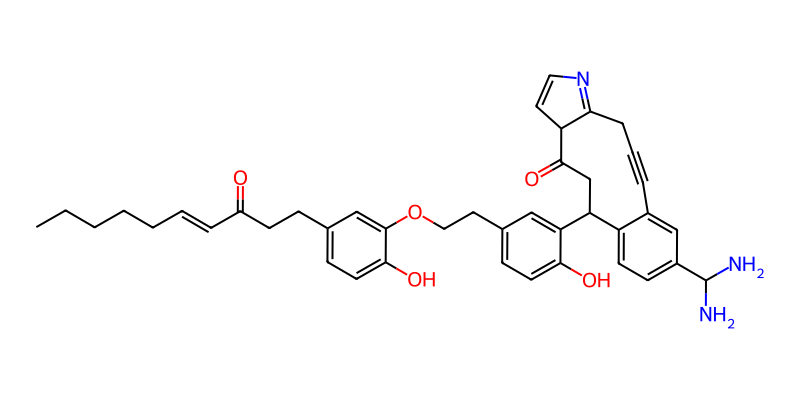}}
\end{minipage} & $-10.63$ & 1.797 & 5.941 \\ \midrule
3 & \begin{minipage}{8em}
      \centering
      \scalebox{0.11}{\includegraphics{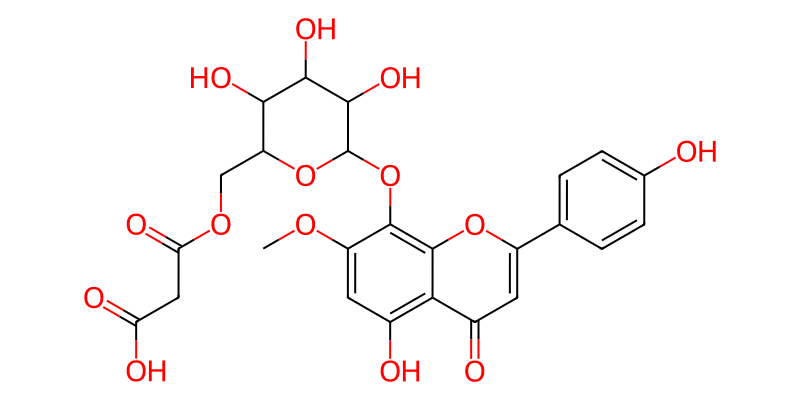}}
\end{minipage} & \begin{minipage}{8em}
      \centering
      \scalebox{0.11}{\includegraphics{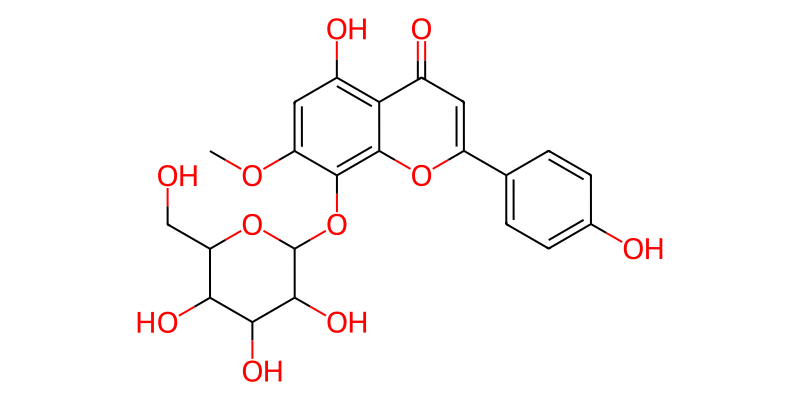}}
\end{minipage} & $-10.28$ & 1.804 & 4.118 \\ \midrule
4 & \begin{minipage}{8em}
      \centering
      \scalebox{0.11}{\includegraphics{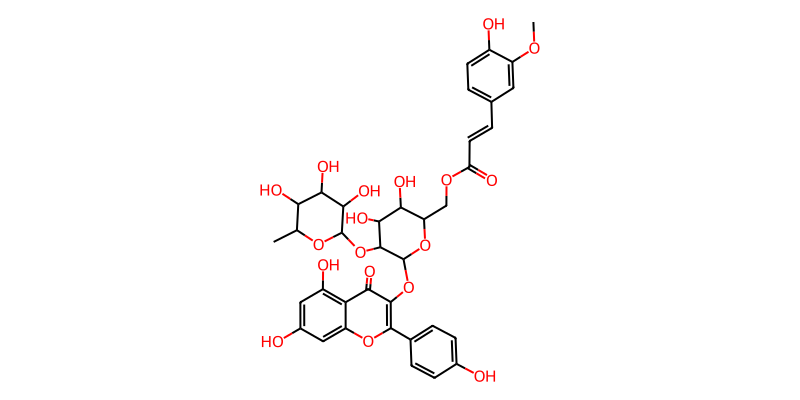}}
\end{minipage} & \begin{minipage}{8em}
      \centering
      \scalebox{0.11}{\includegraphics{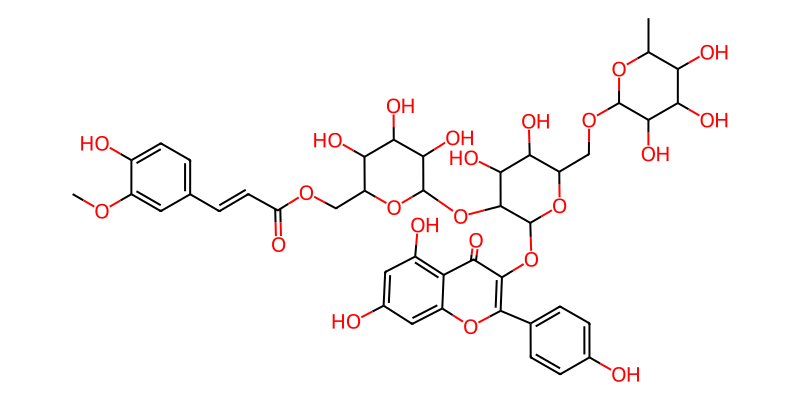}}
\end{minipage} & $-9.99$ & 1.834 & 5.018 \\ \midrule
5 & \begin{minipage}{8em}
      \centering
      \scalebox{0.11}{\includegraphics{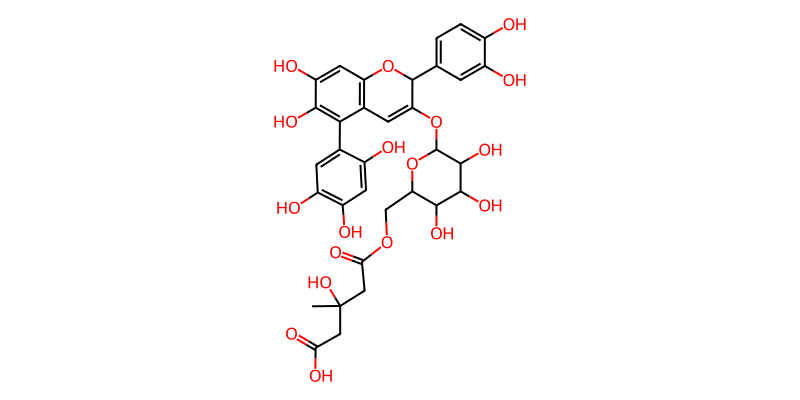}}
\end{minipage} & \begin{minipage}{8em}
      \centering
      \scalebox{0.11}{\includegraphics{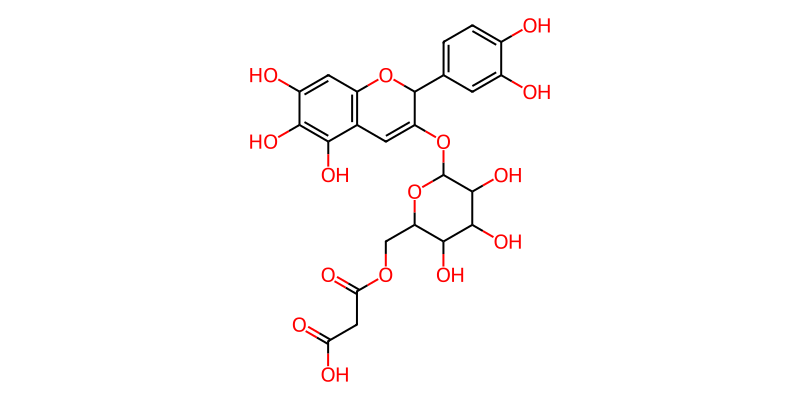}}
\end{minipage} & $-9.96$ & 1.722 & 5.168 \\ \midrule
6 & \begin{minipage}{8em}
      \centering
      \scalebox{0.11}{\includegraphics{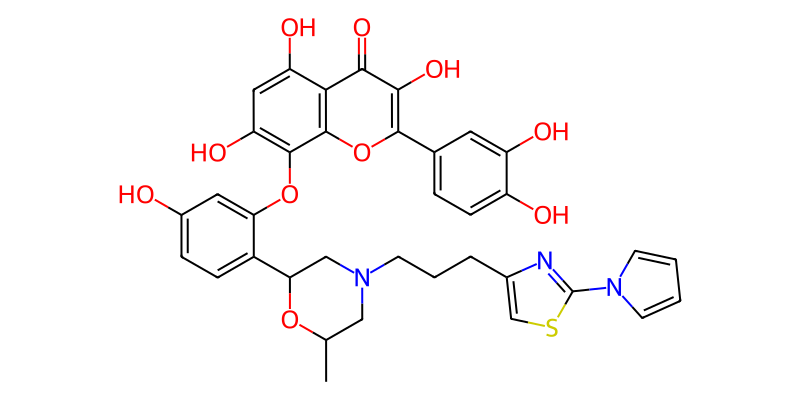}}
\end{minipage} & \begin{minipage}{8em}
      \centering
      \scalebox{0.11}{\includegraphics{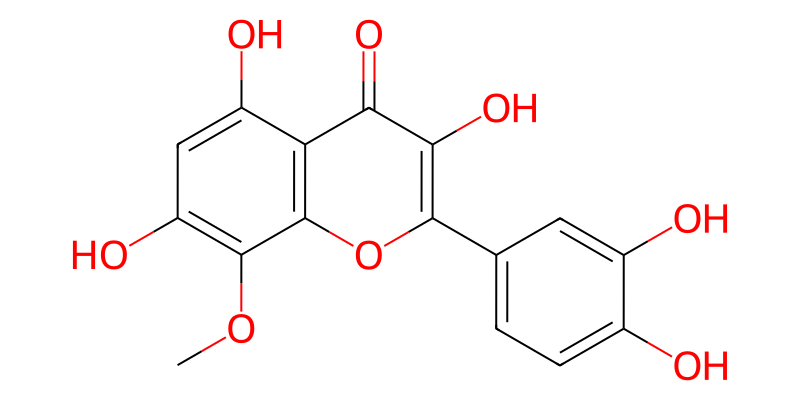}}
\end{minipage} & $-9.96$ & 0.158 & 4.320 \\ \midrule
7 & \begin{minipage}{8em}
      \centering
      \scalebox{0.11}{\includegraphics{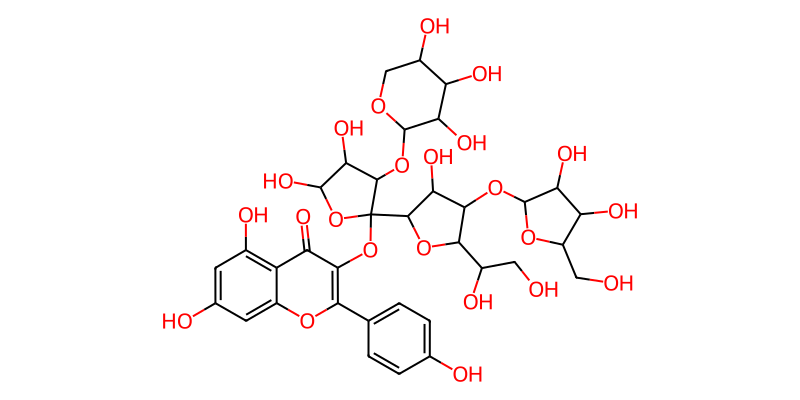}}
\end{minipage} & \begin{minipage}{8em}
      \centering
      \scalebox{0.11}{\includegraphics{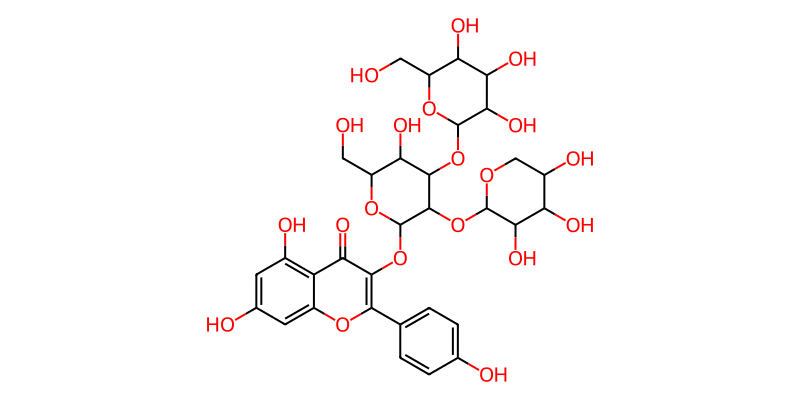}}
\end{minipage} & $-9.96$ & 1.816 & 6.035 \\ \midrule
8 & \begin{minipage}{8em}
      \centering
      \scalebox{0.11}{\includegraphics{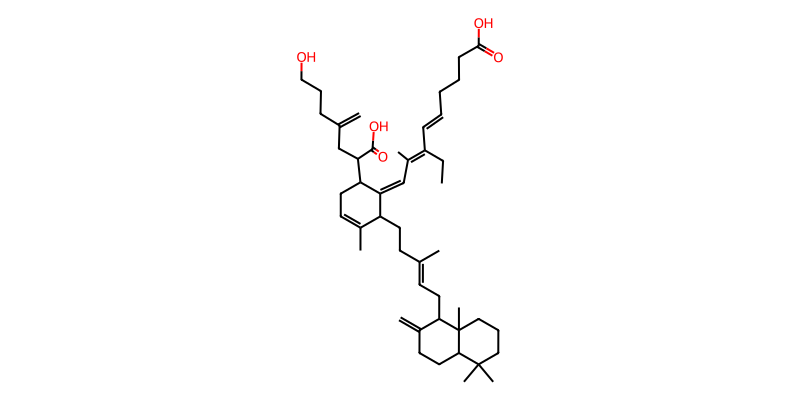}}
\end{minipage} & \begin{minipage}{8em}
      \centering
      \scalebox{0.11}{\includegraphics{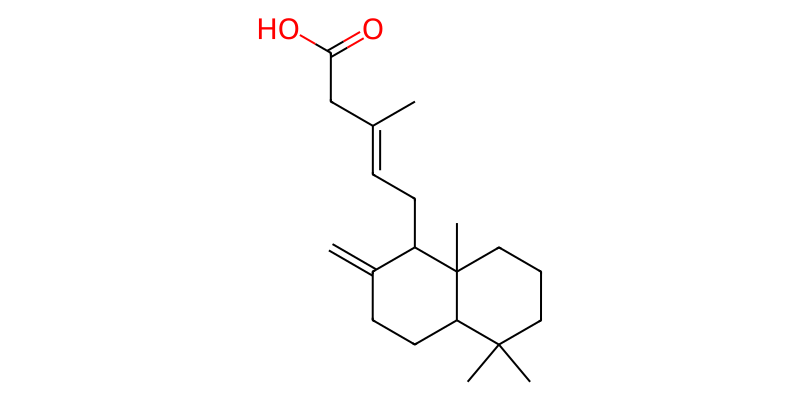}}
\end{minipage} & $-9.96$ & 2.246 & 5.628 \\ \midrule
9 & \begin{minipage}{8em}
      \centering
      \scalebox{0.11}{\includegraphics{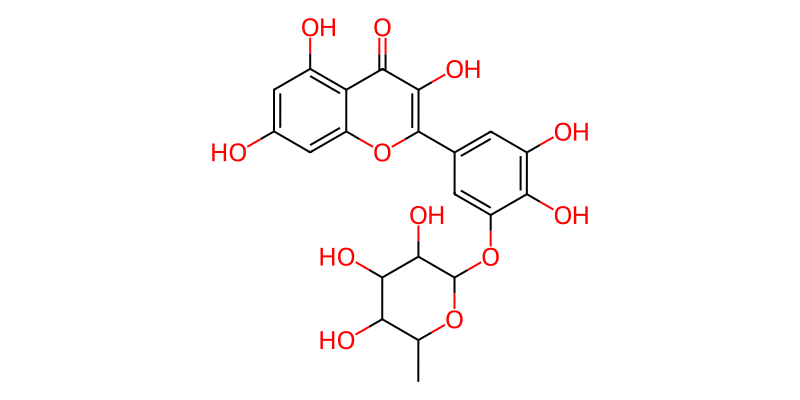}}
\end{minipage} & \begin{minipage}{8em}
      \centering
      \scalebox{0.11}{\includegraphics{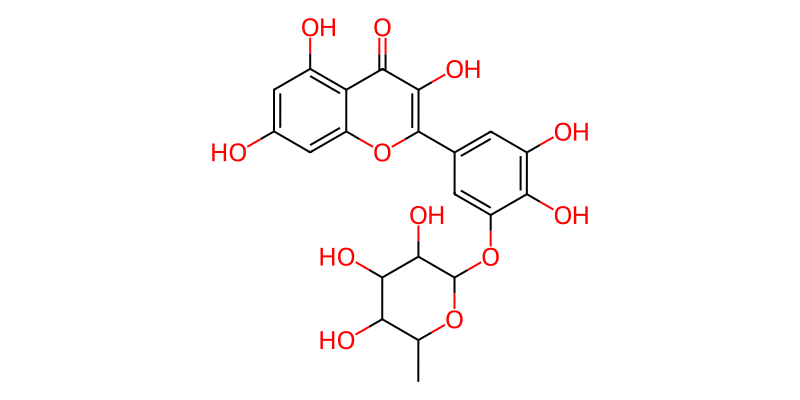}}
\end{minipage} & $-9.96$ & 2.314 & 4.172 \\ \midrule
10 & \begin{minipage}{8em}
      \centering
      \scalebox{0.11}{\includegraphics{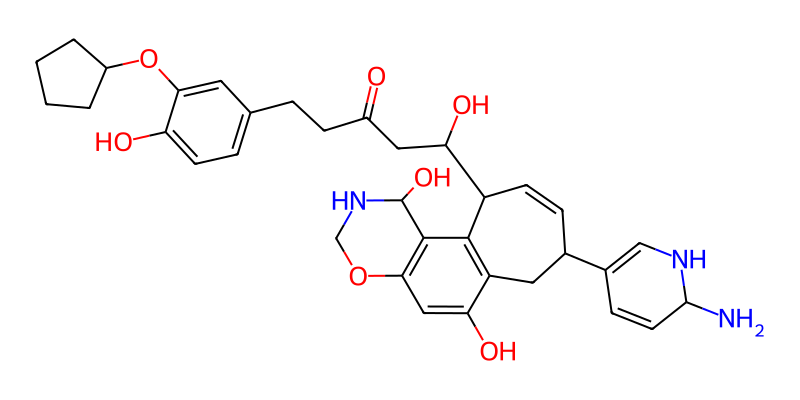}}
\end{minipage} & \begin{minipage}{8em}
      \centering
      \scalebox{0.11}{\includegraphics{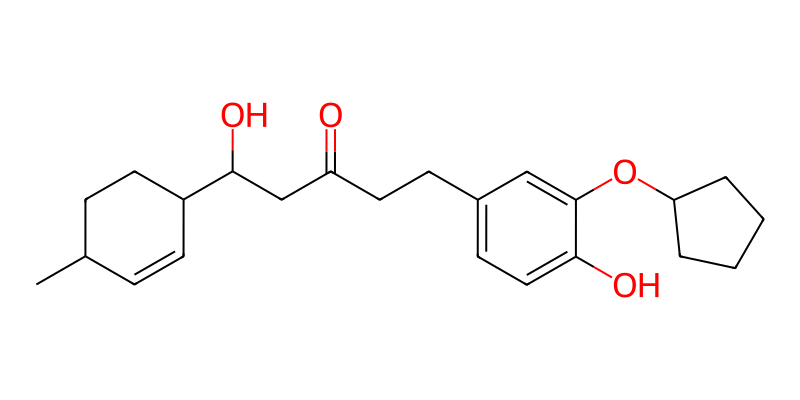}}
\end{minipage} & $-9.96$ & 1.453 & 5.318 \\ \bottomrule
\end{tabular}
\end{table}

\section{Conclusion}
In this research, we fine-tuned a language model pretrained on a natural product dataset to generate natural product-like compounds. We measured various metrics for the molecules generated by the fine-tuned model and demonstrated that they are closer to the distribution of natural products.

In the docking experiments with EGFR, we found that the molecules generated by the fine-tuned smiles-gpt model included viable drug candidates. This illustrates the effectiveness of the language model developed in this research in creating a collection of potential pharmaceutical candidate compounds.

Compared to the previous research by Tay \etal~\cite{Tay2023-nr}, we have been able to create a model that generates compounds that are closer to natural products. Furthermore, this study demonstrates the relationship between the similarity of natural products and the potential utility as drug candidates, which distinguishes it from the previous study. Moreover, there is a need to develop methodologies to extract knowledge from functional structures, such as the potential bioactivity of natural products. Visualization studies focusing on substructures \cite{kengkanna2024, wu2023} may prove to be a valuable tool in this area.

\section*{Declarations}
\subsection*{Funding}
This work was financially supported by the Japan Science and Technology Agency FOREST (Grant No. JPMJFR216J), the Japan Society for the Promotion of Science KAKENHI (Grant Nos. JP23H04880 and JP23H04887), and the Japan Agency for Medical Research and Development Basis for Supporting Innovative Drug Discovery and Life Science Research (Grant No. JP24ama121026).

\subsection*{Conflict of interest}
The authors have no conflicts of interest to declare.

\subsection*{Ethics approval}
Not applicable.

\subsection*{Consent to participate}
Not applicable.

\subsection*{Consent for publication}
Not applicable.

\subsection*{Availability of data and materials}
Not applicable.

\subsection*{Code availability}
NPGPT is available at \url{https://github.com/ohuelab/npgpt} under the MIT license.

\subsection*{Authors' contributions}
KS and MO conceived the idea of the study. KS developed the computational methods and conducted implementation and analyses. KF and MO contributed to the interpretation of the results. KS drafted the original manuscript. MO supervised the conduct of this study. All authors reviewed the manuscript draft and revised it critically on intellectual content. All authors approved the final version of the manuscript to be published.

\bibliography{sn-bibliography}

\end{document}